\documentclass[%
 reprint,
superscriptaddress,
 amsmath,amssymb,
 aps,
]{revtex4-2}

\usepackage{graphicx}
\usepackage{dcolumn}
\usepackage{bm}
\usepackage{braket}

\usepackage[table]{xcolor}
\usepackage{float}
\usepackage{siunitx}
\usepackage{amsmath}
\usepackage{xcolor}

\begin{document}

\preprint{APS/123-QED}


\title{Unitary Design of Quantum Spin Networks for Robust Routing, Entanglement Generation, and Phase Sensing}
 

\author{Abdulsalam H. Alsulami}%
\email{aha555@york.ac.uk}
 \affiliation{%
 York Centre for Quantum Technologies, Department of Physics, University of York, York, YO105DD
 }%
\author{Irene D'Amico}%
\email{irene.damico@york.ac.uk}
 \affiliation{%
 York Centre for Quantum Technologies, Department of Physics, University of York, York, YO105DD
 }%
\author{Marta P. Estarellas}%
\email{mpestarellas@qilimanjaro.tech}
 \affiliation{%
 Qilimanjaro Quantum Tech, Barcelona 08007, Spain
 }%
\author{Timothy P. Spiller}%
\email{timothy.spiller@york.ac.uk}
 \affiliation{%
 York Centre for Quantum Technologies, Department of Physics, University of York, York, YO105DD
 }%

\date{\today}

\begin{abstract}
Spin chains can be used to describe a wide range of platforms for quantum
computation and quantum information. They enable the understanding,
demonstration, and modeling of numerous useful phenomena, such as high
fidelity transfer of quantum states, creation and distribution of entanglement,
and creation of resources for measurement-based quantum processing. In
this paper, a more complex spin system, a 2D spin network (SN) engineered
by applying suitable unitaries to two uncoupled spin chains, is studied.
Considering only the single-excitation subspace of the SN, it is demonstrated
that the system can be operated as a router, directing information through the
SN. It is also shown that it can serve to generate maximally entangled states
between two sites. Furthermore, it is illustrated that this SN system can be
used as a sensor device able to determine an unknown phase applied to a
system spin. A detailed modeling investigation of the effects of static disorder
in the system shows that this system is robust against different types
of disorder. 
\end{abstract}
\maketitle


\section{\label{sec:intro}Introduction}
Spin chains provide a generic model for the behaviour of a range of useful tasks in quantum information processing \cite{bose2003quantum}. They can be used to transfer information through the chain \cite{bose2003quantum,bose2007quantum,nikolopoulos2004electron} or to create and distribute entanglement \cite{yung2005perfect,venuti2006long,d2007freezing,estarellas2017robust,wilkinson2017rapid,riegelmeyer2021generation}. Unlike gate-based schemes,
spin chains do not necessarily require the switching on and
off of couplings between qubits. A desirable process may be
achieved through exploiting the natural dynamics of the system
with a fixed Hamiltonian \cite{benjamin2003quantum,zhou2002quantum}. Furthermore, spin chain systems
can describe and be used to model a wide variety of different
physical implementations. Examples include quantum dots \cite{loss1998quantum,nikolopoulos2004electron,d2006quantum}, trapped ions \cite{brown2016co}, superconducting qubits \cite{berggren2004quantum,KaramlouAmirH2022Qtal} and coupled optical waveguides \cite{blanco2016topological}. 

Spin Networks (SN), on the other hand, can have a topology more complex than linear spin chains \cite{ChristandlMatthias2005Ptoa,KayAlastair2011Bopc} and have applications which include quantum sensing \cite{degen2017quantum,giovannetti2011advances,liu2021distributed,guo2020distributed}. Simple examples of SN systems can be generated by connecting spin chains. If this is effected via suitably chosen unitary transformations, clearly the spectrum is preserved and so the physical behaviour of the SN can be engineered by design. The SN that we investigate here is designed via the application of a Hadamard-like unitary on two uncoupled spin chains. 

In this paper, we show that the SN system
we explore can be used as a router
that sends quantum information through
the network according to an external control.
In a perfect SN, the information can
be directed through the network to a chosen site with unit fidelity,
also known as perfect state transfer (PST). We also demonstrate
that the SN system can be used to create a maximally entangled
state between two sites. Furthermore, we will introduce a protocol
for using our SN as a sensor device that retrieves any unknown
phase applied to a particular spin in the system. Finally, in order
for our results to be applied to realistic situations, we will investigate
in detail the effect of different types of static disorder in
the system.

\section{The Model}
\label{model sec}
A linear spin chain system with nearest-neighbor interaction can
be described by the time-independent XY-Hamiltonian as follows:
\begin{equation}      
\label{Eq:SC Hamiltonian}
\mathcal{H}_{xy} = \frac{1}{2} \sum_{i=1}^{N-1} J_{i,i+1} (\sigma_i^x \sigma_{i+1}^x + \sigma_i^y \sigma_{i+1}^y ) 
+ \sum_{i=1}^{N} \frac{\epsilon_i}{2} (\sigma_i^z) 
\end{equation} 
where $N$ is the total number of sites, $J_{i,i+1}$ is the nearest-neighbor
interaction between sites $i$ and $i+1$, and $ \sigma^x_i $, $ \sigma^y_i $, $ \sigma^z_i $ are Pauli operators that represent the spin components for site $i$. In the second term, in Eq.~(\ref{Eq:SC Hamiltonian}), each $\epsilon_i$ represents the on-site energy for an excitation at site $i$. We will generally consider situations where the
on-site energy is independent of the site $i$ and so we can set $\epsilon_i=0$.
However, when we investigate the effect of diagonal disorder in
the system, $\epsilon_i$ must be considered.

In a spin chain prepared to have spin down $\ket{0}$ at all sites, a single-excitation at a site $i$ is defined as a spin up $\ket{r_i}=\ket{00\dots1_i00\dots}$. The number of excitations $ \mathcal{N} = \sum_{i=1}^{N} \frac{1}{2} (\sigma_i^z + \mathcal{I}_{i}) $, where $\mathcal{I}$ is the relevant identity, is conserved. As $\mathcal{N}$ commutes with $\mathcal{H}_{xy}$ even in the presence of disorder, the different excitation-number subspaces decouple. In our calculations, we will only consider the single-excitation subspace, as this suffices to achieve the desired phenomena.   

We initially consider trimer chains where the coupling parameters are constant, so  $J_{i,i+1}=J$. The advantage of such trimer chains is that information can be transferred through the system with PST. Extension to longer spin chains can be achieved by using different PST schemes (e.g. tuning the boundary couplings, $J_{1,2}$  and $J_{N-1,N}$ \cite{wojcik2005unmodulated,oh2012effect,banchi2010optimal,banchi2011nonperturbative} or controlling each coupling parameter, $J_{i,i+1}$ \cite{kay2010perfect,karbach2005spin,kostak2007perfect, christandl2004perfect,nikolopoulos2004electron,christandl2005perfect}). 

To test how well a quantum process is performed, we use fidelity, which calculates the overlap of an evolved initial state $\ket{\psi(0)}$ with a desirable state $\ket{\psi_{des}}$ at a later time $t$
\begin{equation}
\label{equationFidelity Eq}
F(t) = |\bra{\psi_{des}}e^{-i\mathcal{H}t}\ket{\psi(0)}|^2,
\end{equation}
for the relevant static system Hamiltonian $\mathcal{H}$ and with the reduced Plank constant set to be $\hbar=1$.
For the trimer chain, an example of PST is achieved when the fidelity of an evolved initial state $\ket{r_1}=\ket{100}$ against the desirable state $\ket{r_3}=\ket{001}$ is equal to 1. This happens at time $t_m=\pi/(\sqrt{2}J)$. Since at this time any initial state evolves to its reflection about the mid point of the chain, this is called the mirroring time $t_m$ \cite{christandl2005perfect}. In the presence of errors, the mirroring time and the fidelity change, depending on the type of error, as will be discussed below.

In order to quantify how much a quantum state is entangled we use the Entanglement of Formation (EOF), which gives the entanglement of any arbitrary pair of qubits $A$ and $B$ independent of whether they are in a pure or mixed state. The EOF is defined as \cite{wootters2001entanglement}
\begin{equation}
    EOF_{AB} = -x\log_2x-(1-x)\log_2(1-x) 
\end{equation} 
where $x=\frac{1+\sqrt{1-\tau}}{2}$, $\tau=[max(\lambda_1-\lambda_2-\lambda_3-\lambda_4,0)]^2$, $\lambda_i=\sqrt{\varepsilon_i}$, and $\varepsilon_i$ are the eigenvalues of the matrix $\rho_{AB}\overline{\rho_{AB}}$. Here $\rho_{AB}$ is the reduced density matrix of sites $A$ and $B$, and $\overline{\rho_{AB}}$ is the spin-flipped $\rho_{AB}$, so $\overline{\rho_{AB}}=(\sigma_y^A\otimes\sigma_y^B)\rho^*_{AB}(\sigma_y^A\otimes\sigma_y^B)$.

\subsection{Off-Diagonal Disorder}
Off-diagonal disorder (also called coupling disorder) affects the system coupling parameters. We investigate the effect of the coupling errors on the system by adding these to the coupling parameters of the Hamiltonian $\mathcal{H}_{xy}$ in Eq.~(\ref{Eq:SC Hamiltonian}) 
\begin{equation} 
\label{equationoff-diag error} 
J_{i,i+1}^{perturbed} = J_{i,i+1} + J_{i,i+1}^{'} 
\end{equation}
where $J_{i,i+1}^{'} = E d_{i,i+1}$; the parameter $E$, reported in the figures in units of $\max \{J_{i,i+1}\}$, sets the scale of the error; and $d_{i,i+1}$ is a random number that depends on a particular distribution. Two physically reasonable distributions (uniform and Gaussian distribution) will be considered, each with zero mean value. 

The normalised uniform (or flat) distribution of random numbers is chosen to be of unit weight within the window [-0.5\;,\;0.5]. For zero mean, the normalised Gaussian distribution with a standard deviation of $w$ takes the form 
$f(d) = \frac{1}{w \sqrt{2 \pi}} \exp\left(-d^{2}/2 w^{2} \right)$. As the standard deviation of our chosen flat distribution is $w = \frac{1}{2 \sqrt{3}}$, we use this value of $w$ in the Gaussian distribution to model Gaussian errors equivalent to the flat case. As will be seen, our disorder modelling is then essentially independent of the {\it form} of the random distribution used (flat or Gaussian), until the regime of very large disorder is reached. For error regimes of interest for useful devices, no dependence on the form of the error distribution used will be apparent.

\subsection{Diagonal Disorder}
The second type of disorder considered here is the diagonal disorder
(i.e., on-site energy disorder). Therefore, the second term
in Eq.~(\ref{Eq:SC Hamiltonian}) is now present. The site-dependent errors are represented as $\epsilon_i = Ed_{i}$, where $E$ and the random $d_{i}$ have the same definitions as before.

\section{Spin Network}
\label{SN sec}
Unitary transformations can play a very useful role in designing
and engineering new physical SN systems. Such transformations
can be used to simplify complex networks, while retaining the
spectrum and desirable features of the system dynamics \cite{riegelmeyer2021generation}. Correspondingly, here we use such transformations to couple together
smaller SNs into larger systems, again retaining or engineering
in desirable features. We stress that this unitary transformation
approach is a mathematical design stage, so the resultant SN is
then the network that should be constructed physically, in order
to deliver the desirable dynamics or properties.

In this work, our proposed SN is a 2D system of spins that
is realized by connecting two identical spin chains. In fact, in
general, multiple spin chains can be connected to form a larger
spin network system. However, the applications we study here
only require a small SN, realized by connecting just two identical
spin chains.

\subsection{Design and Realisation of Our SN} 
\label{Realisation of SN}
\begin{figure}[ht]
    \centering
    \includegraphics{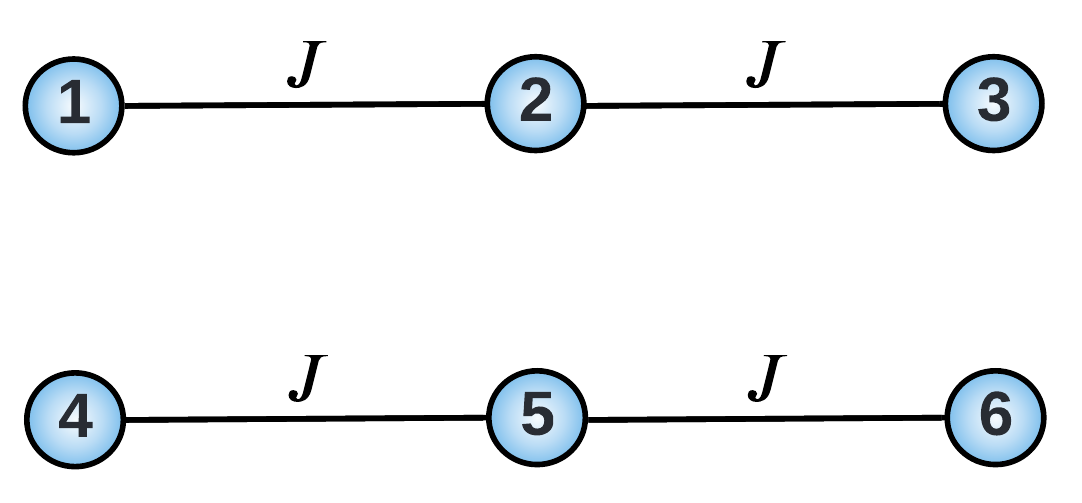}
    \caption{Two uncoupled trimers.  } 
    \label{fig:two uncoupled SC}  
\end{figure}

First, we consider two identical uncoupled trimers as illustrated in Fig.~\ref{fig:two uncoupled SC}. Each trimer acts as a separate spin chain and exhibits PST for a single excitation transferring between site 1 and site 3; similarly, for the second chain, PST is achieved between sites 4 and 6. 

The Hamiltonian for these two uncoupled chains is given as 
\begin{equation}
\label{equationuncoupled trimers} 
H = 
    \begin{pmatrix}  
    0&J&0&0&0&0 \\
    J&0&J&0&0&0 \\
    0&J&0&0&0&0 \\
    0&0&0&0&J&0 \\
    0&0&0&J&0&J \\
    0&0&0&0&J&0  
\end{pmatrix}
\end{equation} \;  \\ 

The two uncoupled spin chains can be connected in order to
form a SN by applying an appropriate unitary transformation $U$
to the Hamiltonian $H$. For the example presented here, we apply
a unitary that is a Hadamard transformation between site 3 and
site 6:
\begin{equation}
\label{equationHadamard-like Unitary}
U =  
    \begin{pmatrix}  
    1&0&0&0&0&0 \\
    0&1&0&0&0&0 \\
    0&0&\frac{1}{\sqrt2}&0&0&\frac{1}{\sqrt2} \\
    0&0&0&1&0&0 \\
    0&0&0&0&1&0 \\
    0&0&\frac{1}{\sqrt2}&0&0&-\frac{1}{\sqrt2}  
\end{pmatrix}
\end{equation} \;
Here, $U^{-1}=U$, and therefore the transformed Hamiltonian is given as 
\begin{equation}
\label{equationSN H} 
\begin{split}
 &\mathcal{H} = UHU^{-1} = \begin{pmatrix}       
    0&J&0&0&0&0 \\
    J&0&\frac{J}{\sqrt{2}}&0&0&\frac{J}{\sqrt{2}}\\
    0&\frac{J}{\sqrt{2}}&0&0&\frac{J}{\sqrt{2}}&0 \\
    0&0&0&0&J&0 \\
    0&0&\frac{J}{\sqrt{2}}&J&0&-\frac{J}{\sqrt{2}} \\
    0&\frac{J}{\sqrt{2}}&0&0&-\frac{J}{\sqrt{2}}&0   
\end{pmatrix}
\end{split}
\end{equation} \;

\begin{figure*}
\includegraphics{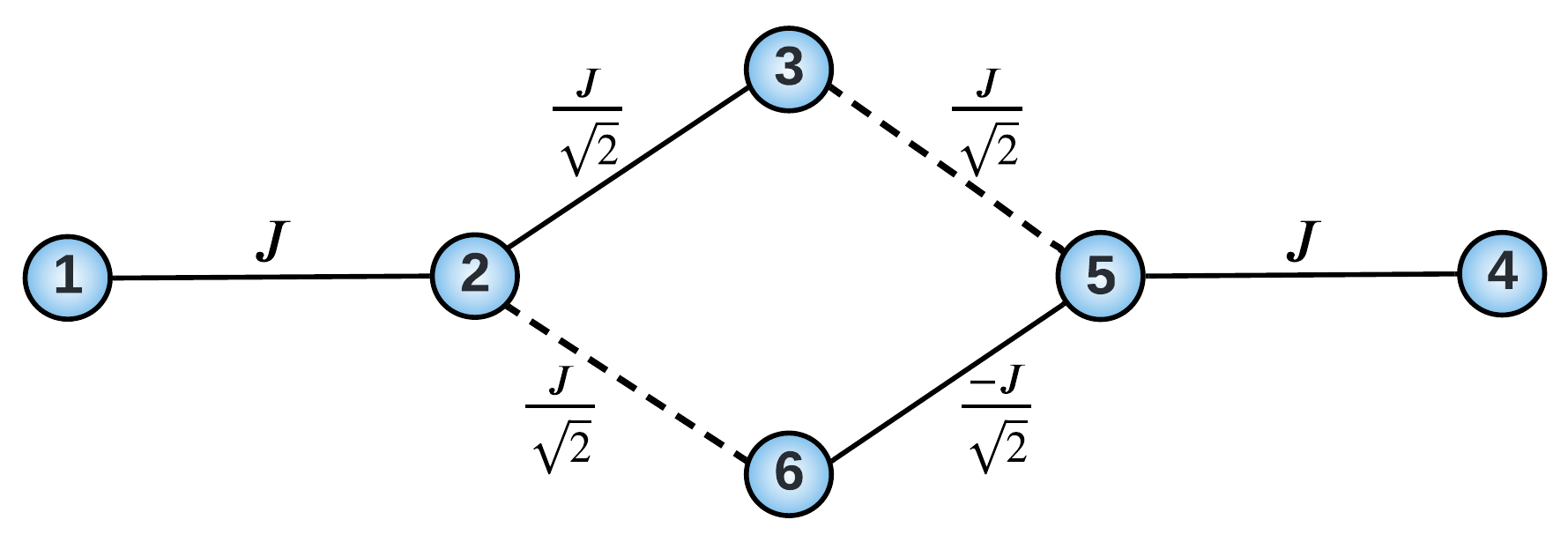}
\caption{Scheme of our 6 sites SN system. The dashed lines are the new couplings that connect the two uncoupled chains as a result of the unitary transformation of the Hamiltonian. Note that on top of these two additional couplings the energy associated to some of the already existing ones has changed.}
\label{fig:6SN}      
\end{figure*}

Using this transformation, we have designed a SN system of
six sites (Fig.~\ref{fig:6SN}).We stress that it is this coupled SN that should
be constructed physically, in order to deliver the various applications.
We choose to design the SN via a unitary transformation
because we know that the spectrum is unchanged. Therefore U
is chosen to produce interesting and useful transformed dynamics.
In order to demonstrate this, we require the eigenvectors and
eigenvalues (see appendix A) of the Hamiltonian $\mathcal{H}$. The system is first prepared to be in the state $\ket{000\dots}$ where all sites have spin-down. Then, the dynamics of the system is initialised by the simple injection of a single-excitation at site 1 at $t=0$:
\begin{equation}
\label{equationexcit at site1 SN}   
\ket{\psi_1(0)} = \ket{r_1}, 
\end{equation}
where $\ket{r_1}=\ket{100000}$.

This state is clearly not an eigenstate, so it will evolve in a non-trivial way. The state evolution is then found numerically\footnote{we use the Python library: scipy.linalg.expm($-i\mathcal{H}t$)} or analytically by decomposing the initial state Eq.~(\ref{equationexcit at site1 SN}) into the eigenvectors of the Hamiltonian $\mathcal{H}$. The excitation will then evolve, through the natural dynamics of the system, to a state that is a superposition between site 3 and site 6 at the mirroring time $t_m$

\begin{equation}
\label{equations3&s6} 
\ket{\psi_1(t_m)} = -\frac{1}{\sqrt{2}}(\ket{r_3} + \ket{r_6}).
\end{equation} \; 
Then it evolves back to site 1 at $2t_m$ and so $\ket{\psi_1(0)}=\ket{\psi_1(2t_m)}$. 

Similarly, if at $t=0$ we inject the single-excitation at site 4 instead of site 1, then the excitation will evolve at $t_m$ to a superposition state between site 3 and site 6 but with a relative phase of $-1$ 
\begin{equation}
\label{equation4 to 3&6} 
\ket{\psi_4(t_m)} = -\frac{1}{\sqrt{2}}(\ket{r_3} - \ket{r_6}).
\end{equation} \; 
Then, it evolves back to site 4 at $2t_m$.

This dynamics is engineered by constructing a Hadamard-based unitary to connect the chains; unitary based on different gates could be used for engineering alternative behaviours/state superpositions.   

For demonstration, in what follows, we will use just the case where the single-excitation is injected at site 1. 

\subsection{Router}
Routing is clearly an important function to have in quantum networks. Spin chains have been proposed as routers via time-dependent driving of the chain couplings \cite{bayat2010entanglement,paganelli2013routing}  or modulating the on-site energies \cite{zueco2009quantum}. In our scheme, the Hamiltonian parameters will instead remain unchanged.

We now explain how our SN can be used as a router, so to control whether the excitation is allowed to return to site 1 or to propagate to site 4. We have seen above that the excitation evolves from site 1 to a superposition of being at site 3 and site 6 and then it evolves back to site 1. For routing, we perform a sudden (on the timescale of the dynamics) operation at the time $t_m$, when the excitation arrives at site 3 and site 6, to force the excitation to evolve to site 4. This is done by applying a local phase flip of $(e^{i\pi}=-1)$ at either site 3 or site 6 at $t_m$. For illustration, we choose to apply the phase flip at site 6, such that under the sudden approximation the state at $t_m$ becomes:
\begin{equation} 
\label{equation3& pf_6}      
\ket{\psi_1(t_m)}_{\pi} = -\frac{1}{\sqrt{2}}(\ket{r_3} + e^{i\pi}\ket{r_6}).
\end{equation} \; 
As a result, the evolution of $\ket{\psi_1(t_m)}_{\pi}$ will transfer the excitation to site 4 at $2t_m$. Therefore,  the system will be operated as a router such that the excitation will not evolve back to site 1, but it will evolve to site 4 instead. 

Clearly, by design as it is based on PST, the router will operate
perfectly for a perfect SN. However, in anticipation of considering
practical systems with disorder present, we use fidelity to asses
how well the excitation is transferred to site 4. In order to do so,
we should take into account that the evolution changes when a
phase factor is applied to the system. Consequently, there are two
fidelities to be calculated, (i.e., the fidelity before and after the
phase factor is applied to the system). For instance, if the system
is initialized with a single-excitation injected at site 1 at $t=0$ and we add a phase factor $e^{i\theta}$ at site 6 at $t_m$, the fidelity of the excitation being at the  desirable site $\ket{\psi_{des}}$ is calculated as follows:
\begin{enumerate}
    \item The fidelity for $0\leq t <t_m$ is calculated as: \\ 
    $F(t)=|\bra{\psi_{des}}\exp(-i\mathcal{H}t)\ket{\psi_1(0)}|^2$, with the initial state being $\ket{\psi_1(0)}=\ket{r_1}$.  
    \item The fidelity for $t\geq t_m$ is calculated as: \\ 
    $F(t)=|\bra{\psi_{des}}\exp(-i\mathcal{H}t)\ket{\psi_1(t_m)}_{\theta}|^2$, 
    with the new initial state being $\ket{\psi_1(t_m)}_{\theta} = -\frac{1}{\sqrt{2}}(\ket{r_3} + e^{i\theta}\ket{r_6})$. 
\end{enumerate} 

\begin{figure}[ht!] 
    \centering
    \includegraphics[width = 0.45\textwidth]{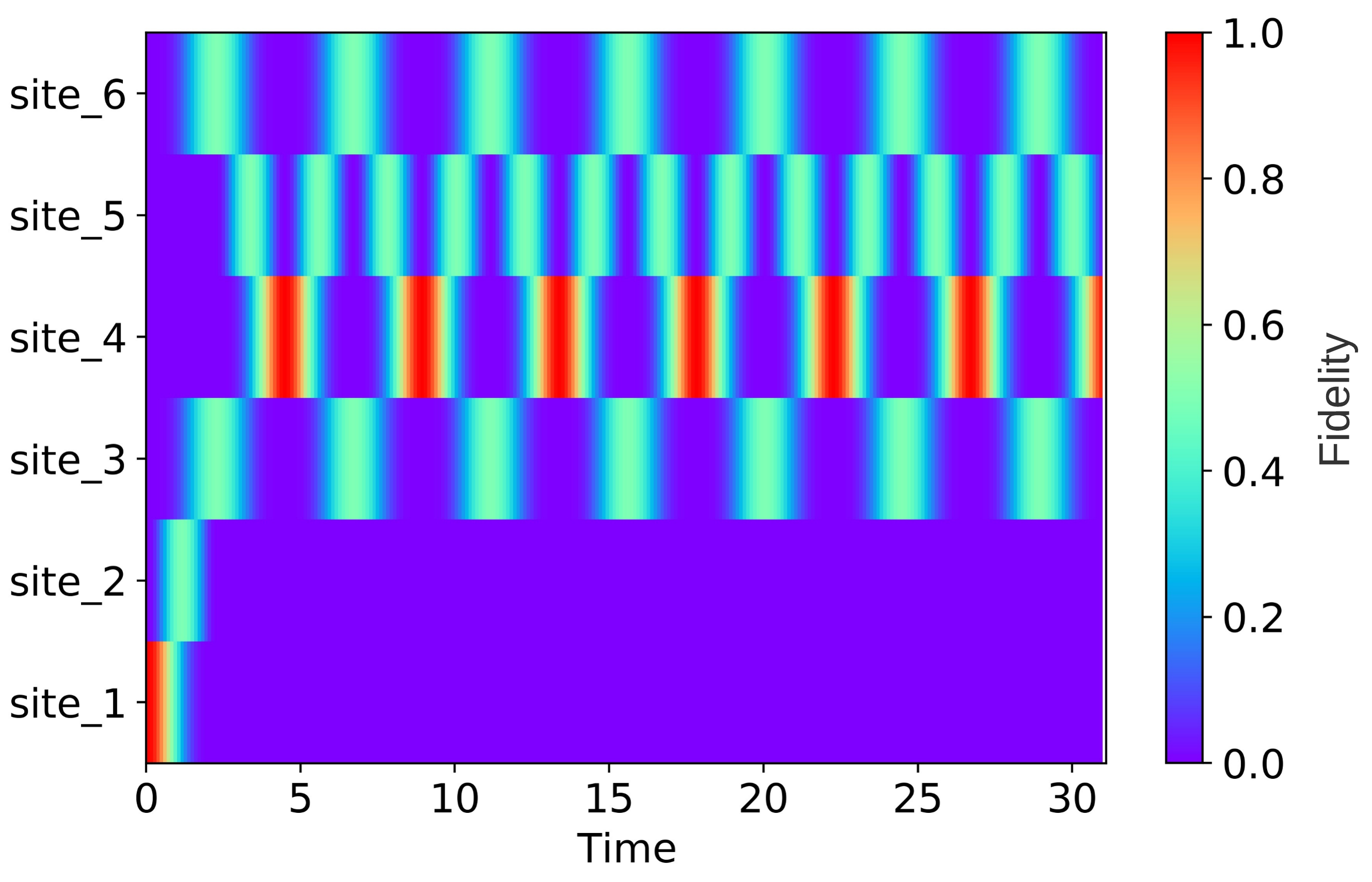} 
    \caption{The fidelity of an excitation being at each of the six sites as a function of time. PST is achieved from site 1 to site 4 when a phase flip is applied at site 6 at $t_m$. The colored bar shows the fidelity values ranging from 0 to 1.}
    \label{fig:site 4 fid} 
\end{figure}

\begin{figure}[ht!] 
    \centering
    \includegraphics[width = 0.45\textwidth]{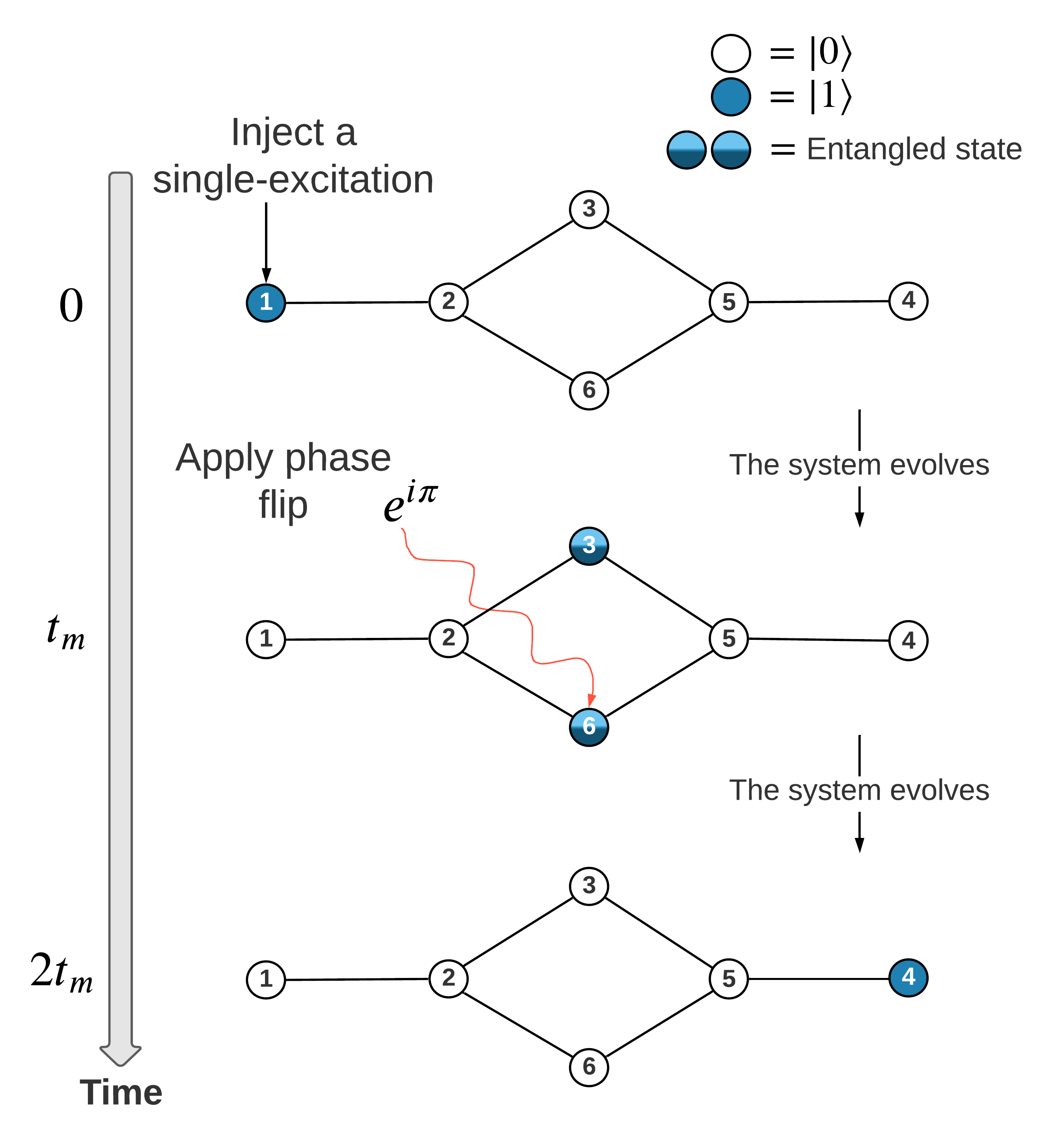}
    \caption{The routing protocol is achieved by injecting a single-excitation at site 1 at $t=0$ and injecting a phase flip at site 6 at $t_m$.}
    \label{fig:6s SN protocol}
\end{figure}

The desirable state for the routing device is an excitation being at site 4, $\ket{\psi_{des}}=\ket{r_4}$. In Fig.~\ref{fig:site 4 fid}, the fidelity of an excitation being at each site is plotted as a function of time, which shows that the excitation is transferred from site 1 to site 4 with PST. The excitation will then keep oscillating between site 4 and a superposition state of site 3 and site 6. If the phase flip is applied repeatedly at site 6, each time the excitation is in a superposition of site 3 and site 6 (i.e., $t_m$, $3t_m$, $5t_m$, \dots), then the excitation will keep oscillating between site 1 and site 4. The protocol for operating the SN as a router is shown in Fig.~\ref{fig:6s SN protocol}.

\subsubsection{Router Robustness} 
\begin{figure}[ht!]
    \centering
    \includegraphics[width = 0.45\textwidth]{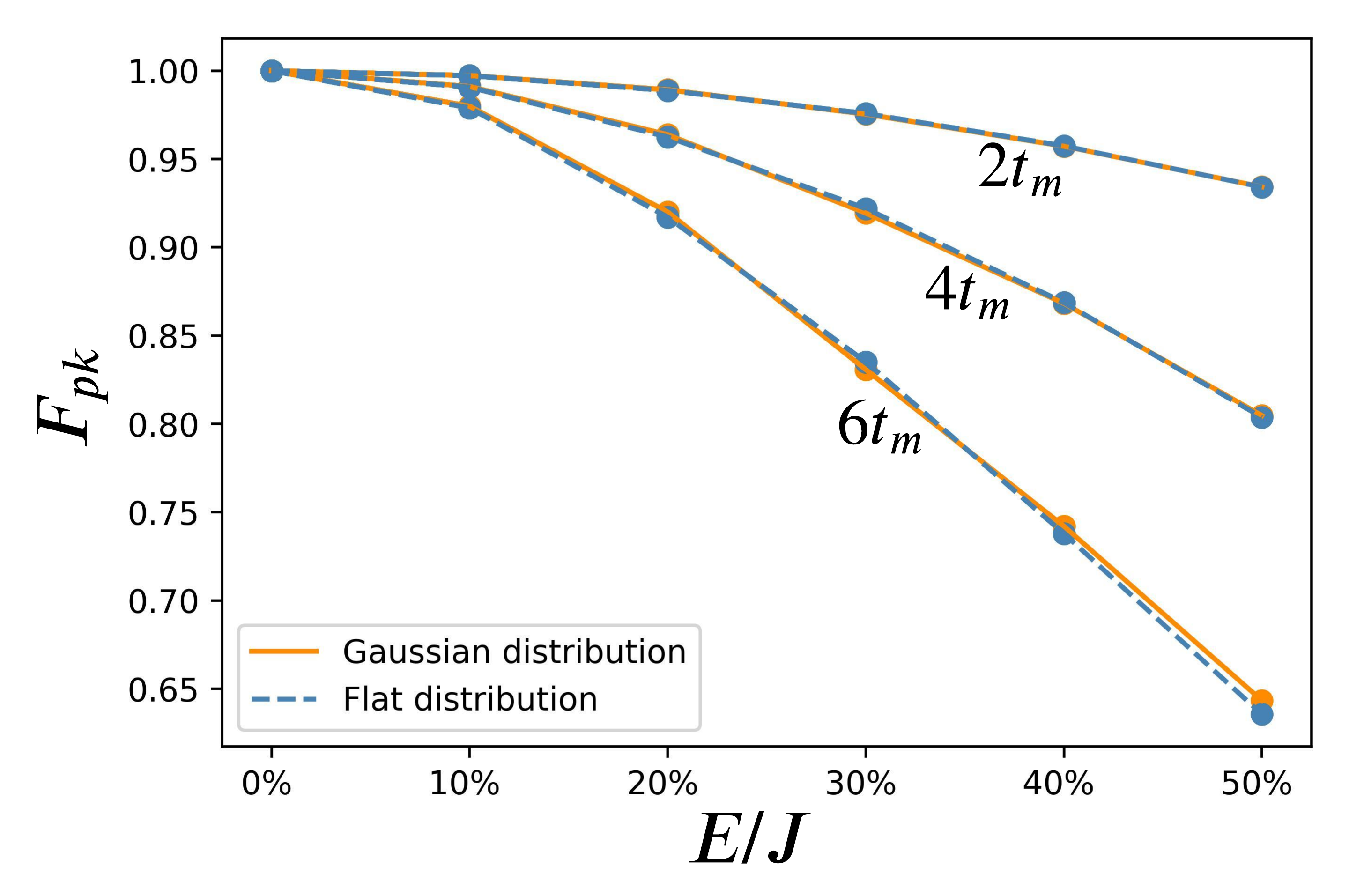}
    \caption{Fidelity peaks $F_{pk}$  ($F(t)$ against the state $\ket{r_4}$ for $t=2t_m$, $4t_m$, and $6t_m$) for the routing protocol in the presence of diagonal disorder with different error scales $E/J$ and for random Gaussian and flat distributions (solid orange and dashed blue line, respectively). Each point has been averaged over 1000 realisations of the fidelity.}
    \label{fig: s4 fid vs diag error}
\end{figure}

\begin{figure}[ht!] 
    \centering
    \includegraphics[width = 0.45\textwidth]{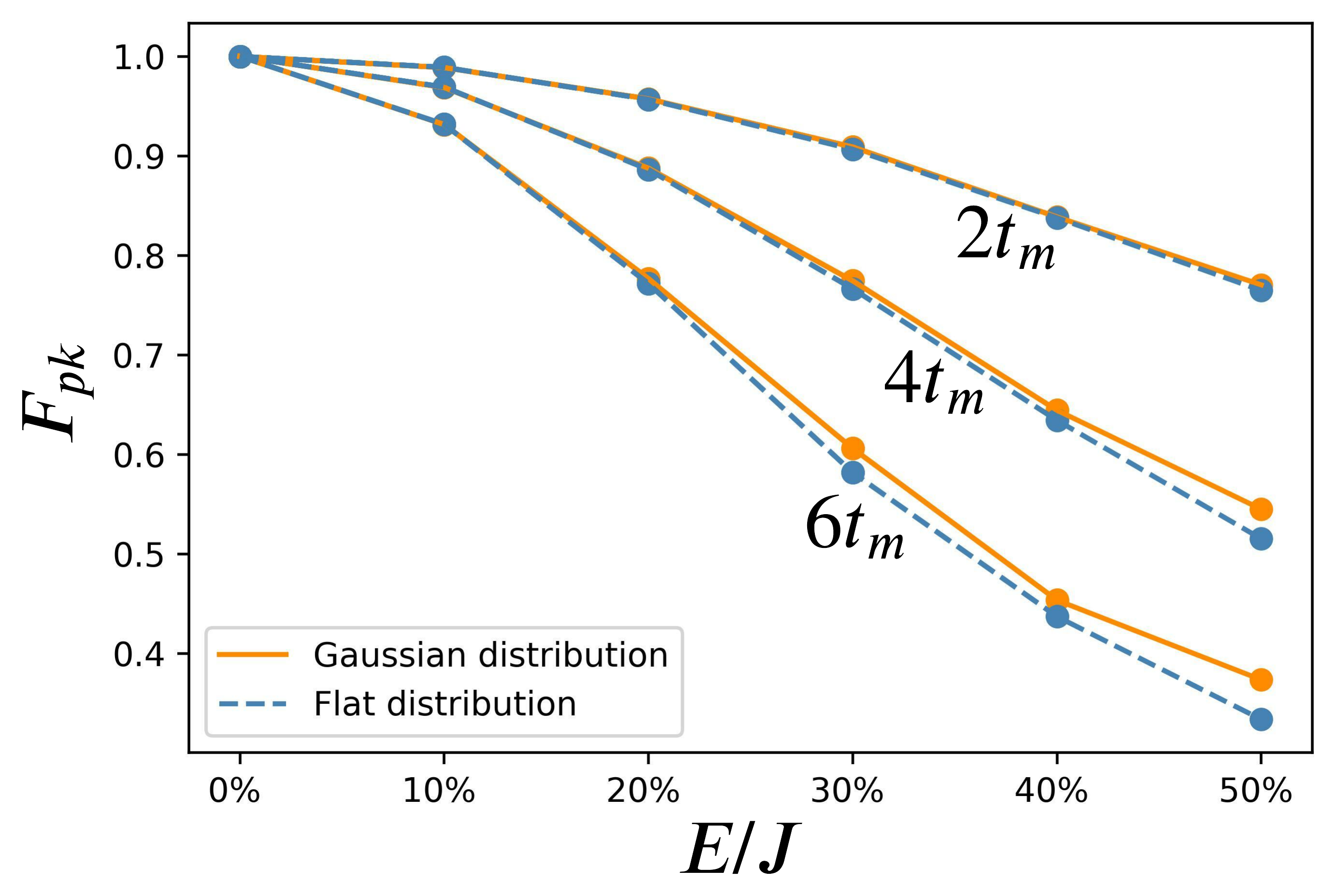} 
    \caption{Fidelity peaks $F_{pk}$  ($F(t)$ against the state $\ket{r_4}$ for $t=2t_m$, $4t_m$, and $6t_m$) for the routing protocol in the presence of off-diagonal disorder with different error scales $E/J$ and for random Gaussian and flat distributions (solid orange and dashed blue line, respectively). Each point has been averaged over 1000 realisations of the fidelity.}
    \label{fig: s4 fid vs off-diag error}
\end{figure}

In order to test the robustness of our routing protocol we apply two types of disorder, diagonal and off-diagonal disorder. We then measure the fidelity against $\ket{r_4}$ at different times $2t_m$, $4t_m$, and $6t_m$. These are the times where the excitation is found to be at site 4 for the ideal case with no errors. For both types of disorder we have considered a Gaussian and a flat distribution for the random  error $d_i$, with error scale $E/J$. Since the error is random, we consider a large number of realisations so each point in Fig.\eqref{fig: s4 fid vs diag error} and Fig.\eqref{fig: s4 fid vs off-diag error} is the average value of 1000 realisations of the fidelity in the presence of disorder.


A very robust behaviour of the fidelity against diagonal disorder can be seen in Fig.~\ref{fig: s4 fid vs diag error} where the fidelity remains above $99\%$ at $2t_m$ and $>95\%$ at a later time, $6t_m$, even with large error scale, $E/J=15\%$. With significant error scale, $E/J=25\%$, the fidelity at $2t_m$, $4t_m$, and $6t_m$ remains above $98\%$, $94\%$, and $88\%$, respectively. On other hand, the fidelity is not so robust against off-diagonal disorder, as shown in Fig.~\ref{fig: s4 fid vs off-diag error}, rapidly decreasing as $E/J\geq20\%$. The plots also show that both random number distributions (Gaussian and flat distributions) give results which are indistinguishable on the scale of the plots for error scale up to $E/J=20\%$, while for $E/J>20\%$ the Gaussian distribution has slightly less impact on the fidelity compared to the flat distribution.  

The routing protocol is very robust for both types of disorder when the error scale is up to $E/J\leq10\%$. Both figures illustrate that when $E/J\leq10\%$, the routing fidelity remains above $97\%$ at three consecutive times ($2t_m$, $4t_m$, and $6t_m$) with diagonal disorder, while it remains above $90\%$ at the same three consecutive times with off-diagonal disorder. It is important to note that in real implementations the error scale could reasonably be expected to be $E/J<10\%$, and since the robustness of our routing protocol remains high for such error scales, we highlight the potential of our SN for short-distance routing applications.

\subsection{Entanglement Generation} 
\label{Entanglement SN}
The SN system initialized as $\ket{r_1}$ naturally generates the maximally entangled state  Eq.~(\ref{equations3&s6}) between sites 3 and 6 at times $nt_m$ with $n$ odd. However, it can also be used to create a maximally entangled state between sites 1 and 4. This is done as follows: We inject a single excitation at site 1 at $t=0$ and we then apply a phase shift of $e^{i\pi/2}$ at site 6 at $t_m$. As a result, the state will evolve at $2t_m$ to be 
\begin{equation}
\label{Entang Eq}
\ket{\psi_1(2t_m)} = \frac{1+e^{i\pi/2}}{2}\ket{r_1}+\frac{1-e^{i\pi/2}}{2}\ket{r_4}. 
\end{equation} \; 

The state of the system then keeps oscillating between a superposition state of site 1 and site 4 and a superposition state of site 3 and site 6, such that the state at $3t_m$ will be

\begin{equation}
\ket{\psi_1(3t_m)}_{\frac{\pi}{2}}= -\frac{1}{\sqrt{2}}(\ket{r_3}+e^{i\pi/2}\ket{r_6})
\label{eq:one}
\end{equation}
and the state at $4t_m$ will return to be 
\begin{equation}
    \ket{\psi_1(4t_m)}= \ket{\psi_1(2t_m)}
\end{equation} \; 

The state $\ket{\psi_1(2t_m)}$ is a maximally entangled state between site 1 and site 4. We note that as the SN could be engineered from longer spin chains, this is an effective way to entangle distant qubits and can be used as a resource for quantum information processing purposes.

\subsubsection{Entanglement Robustness} 
\begin{figure}[ht!]
    \centering
    \includegraphics[width = 0.45\textwidth]{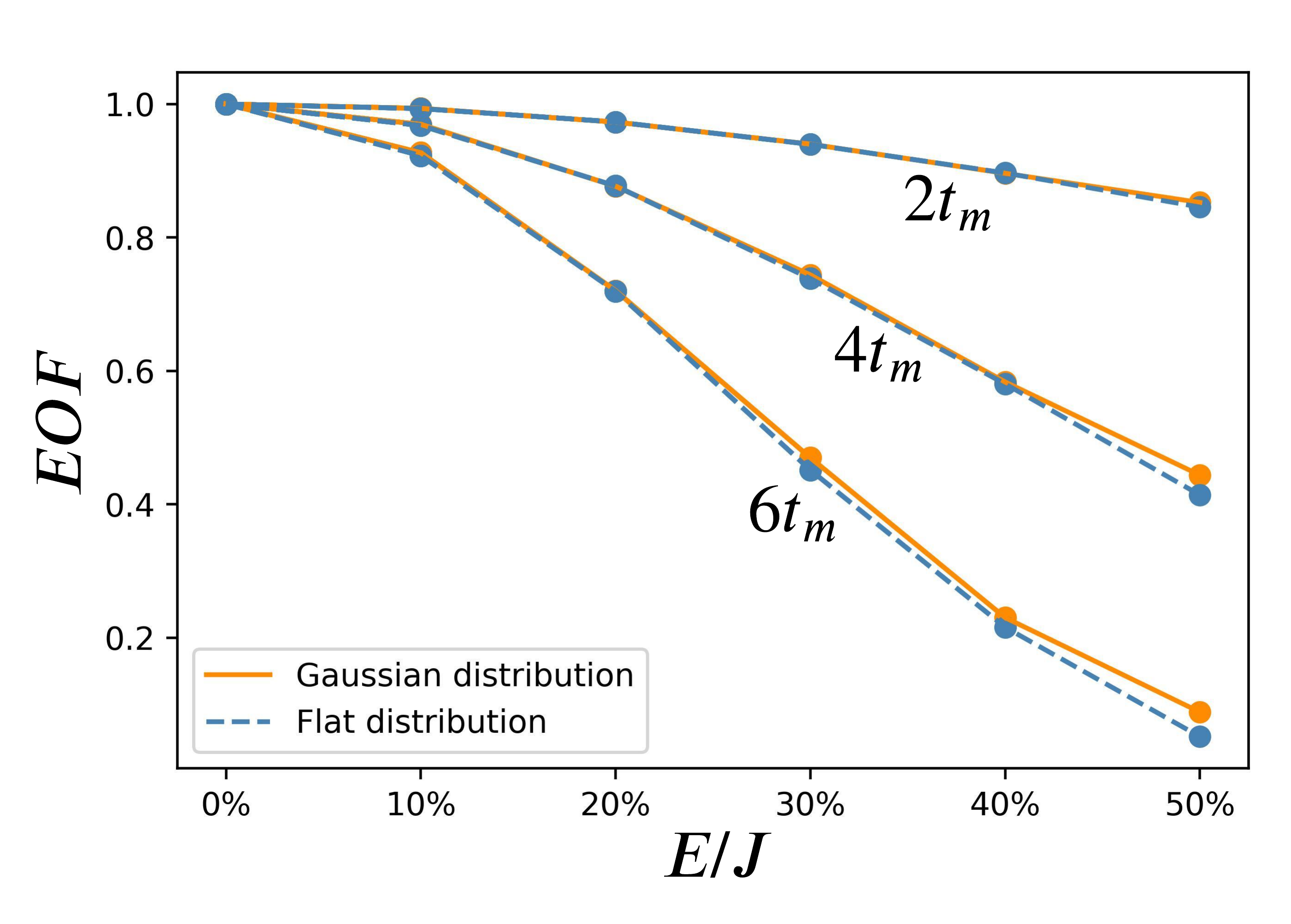}
    \caption{EOF between sites 1 and 4 at $2t_m$, $4t_m$, and $6t_m$ in the presence of diagonal disorder with different error scales $E/J$ and for random Gaussian and flat distributions (solid orange and dashed blue line, respectively). Each point has been averaged over 1000 realisations of the EOF.}     
    \label{fig: EOF vs diag error}
\end{figure}

\begin{figure}[ht!] 
    \centering
    \includegraphics[width = 0.45\textwidth]{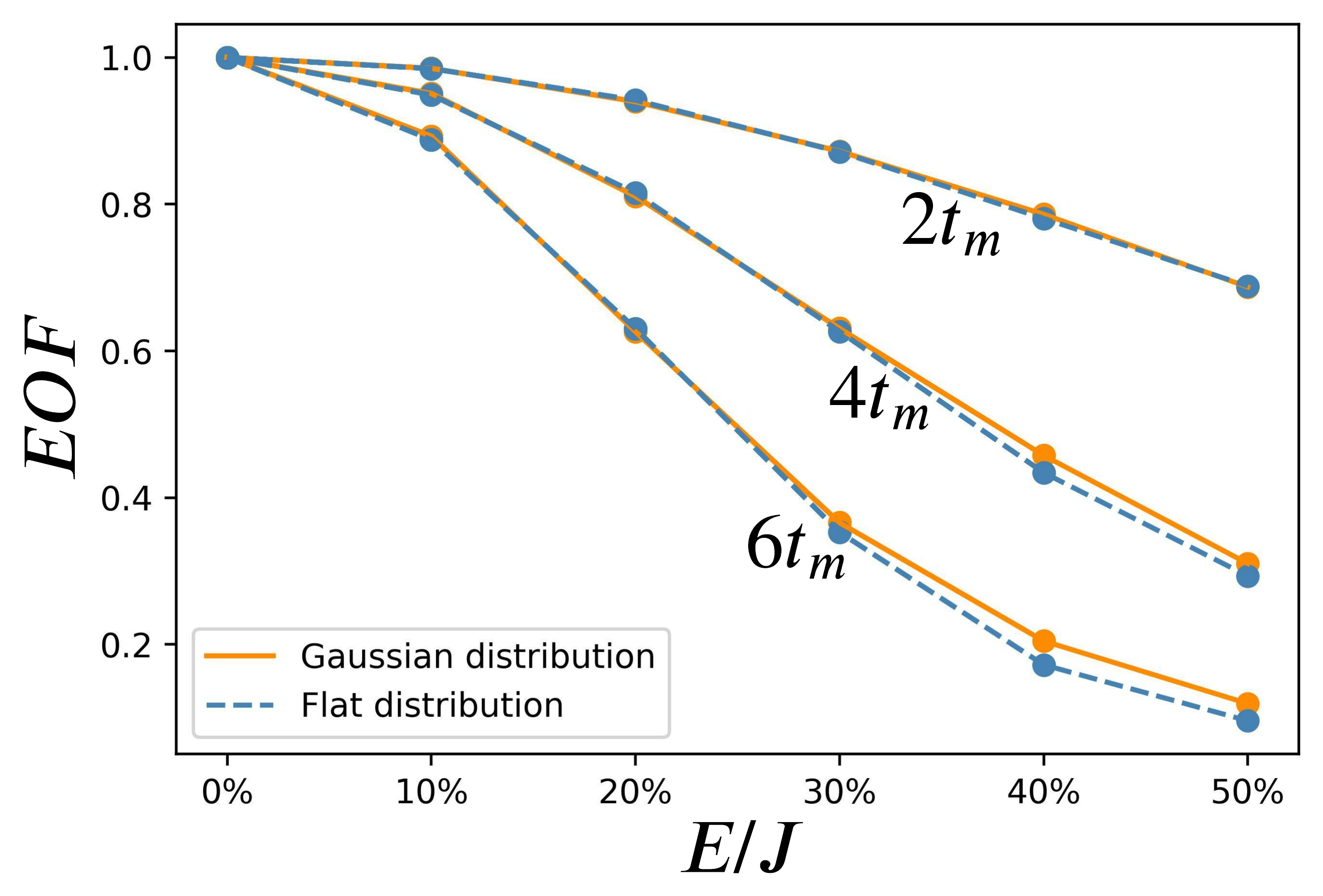} 
    \caption{EOF between sites 1 and 4 at $2t_m$, $4t_m$, and $6t_m$ in the presence of off-diagonal disorder with different error scales $E/J$ and for random Gaussian and flat distributions (solid orange and dashed blue line, respectively). Each point has been averaged over 1000 realisations of the EOF.}     
    \label{fig: EOF vs off-diag error}
\end{figure}

We now wish to investigate the robustness of our protocol to generate an entangled state against error. This is done by calculating the effect of errors on the EOF of the relevant maximally entangled state between site 1 and site 4 at different times $2t_m$, $4t_m$, and $6t_m$. These are the times where the state of the system, for no error, is found to be maximally entangled. Two types of error have been considered, the presence of on-site energy error Fig.~\ref{fig: EOF vs diag error} and  coupling imperfections Fig.~\ref{fig: EOF vs off-diag error}. Since the errors are random, we consider a large number of realisations and each point in Fig.\eqref{fig: EOF vs diag error} and Fig.\eqref{fig: EOF vs off-diag error} has been averaged over 1000 realisations of EOF in the presence of errors.  


We observe that the generation of the entangled state seems to be very robust against diagonal disorder Fig.~\ref{fig: EOF vs diag error}, with an EOF that remains above $99\%$ at $2t_m$ for error scale up to $E/J=15\%$ and above $95\%$ for a significant error scale of $E/J=25\%$. Furthermore, the EOF at times up to $6t_m$ remains above $90\%$ with a large error scale of $E/J=10\%$. In the off-diagonal disorder case, Fig.~\ref{fig: EOF vs off-diag error}, the EOF remains above $97\%$ and $95\%$ when the error scale is $E/J=10\%$ at $2t_m$ and $4t_m$, respectively. However, it decreases fast as $E/J\geq20\%$, particularly at later times ($4t_m$ and $6t_m$).

It is clear for the diagonal disorder case that the effects of the random Gaussian and flat distributions are identical at $2t_m$ up to $E/J=40\%$ and very similar at later times. In the case of off-diagonal disorder, results from both random number distributions remain identical at $2t_m$ and $4t_m$ up to $E/J=30\%$ and extremely close at $6t_m$ up to $E/J=30\%$. Therefore, the EOF behaviour is independent of the form of the random distribution used (Gaussian or flat), until the regime of significant error strength of $E/J\geq30\%$ is reached. The robustness of our protocol for a significant error scale makes it a promising candidate to generate entangled states in real applications.

\subsection{Phase Sensor Device} 
\label{Q sensing sec}
Quantum sensors now form a recognised sector within the whole quantum technologies landscape \cite{degen2017quantum,giovannetti2011advances,liu2021distributed,pirandola2018advances,pezze2021quantum,maleki2022distributed,dorner2009optimal,zhuang2018distributed}. Use of quantum resources for sensing enables the ultimate (Heisenberg-limited) sensitivity to be pursued.  In essentially all such applications, the field or effect being sensed is arranged to impart a phase on the quantum resources (qubits) in some chosen interferometric system. Thus measurement of this phase through interference enables sensing of the field or effect that caused it.

In both implementations mentioned above, a chosen phase is applied at site 6 at $t_m$, to perform a specific operation. We now consider the case where the sudden phase factor $e^{i\theta}$ applied at site 6 is unknown and arises from an external field or effect, which is to be investigated by the phase it generates. The task is to retrieve this unknown phase $\theta$ (modulo $ 2 \pi$), in order to sense the field or effect that produced it. We can derive analytically how an arbitrary phase $\theta$ will modify the occupation of sites 1 and 4 at $2t_m$, see appendix \ref{appendix B}. From a practical perspective, we assume that the only information we could determine is the fidelity measurement at either sites 1 or site 4 against $|r_1\rangle$ or $|r_4\rangle$ respectively, at $2t_m$. In what follows, we will assume we can retrieve the fidelity against $\ket{r_1}$.

When an initial excitation is injected at site 1 at $t = 0$ and an unknown phase $\theta$ is applied suddenly at site 6 at $t_m$, the state of the system at $2t_m$ will be given by         
\begin{equation}   
\label{equation24}   
\ket{\psi_1(2t_m)} = \frac{1+e^{i\theta}}{2}\ket{r_1} +    \frac{1-e^{i\theta}}{2}\ket{r_4}. \; 
\end{equation}     

Thus, by considering the overlap of the initial state $\ket{\psi_1(0)}=|r_1\rangle$  with the system state at $2t_m$, Eq.~(\ref{equation24}), the fidelity against $\ket{r_1}$ at $2t_m$ can be written as
\begin{equation}
\label{F1 eq} 
F_1=\frac{1}{2}(1+\cos{\theta}). \; 
\end{equation} 
More details are given in appendix B.
Therefore, in the ideal case, the unknown angle can be obtained as $\theta=\cos^{-1}({2F_1-1})$. It is important to note that this would enable us to retrieve any unknown angle in the range from $0$ to $\pi$. However, the unknown angle could be in the range from $0$ to $2\pi$. Thus, we also need a separate fidelity as a function of $\sin \theta$, alongside $F_1$, in order to be able to obtain any unknown angle in the range from $0$ to $2\pi$. 

\begin{figure*}
\includegraphics[width=17cm, height=10.5cm]{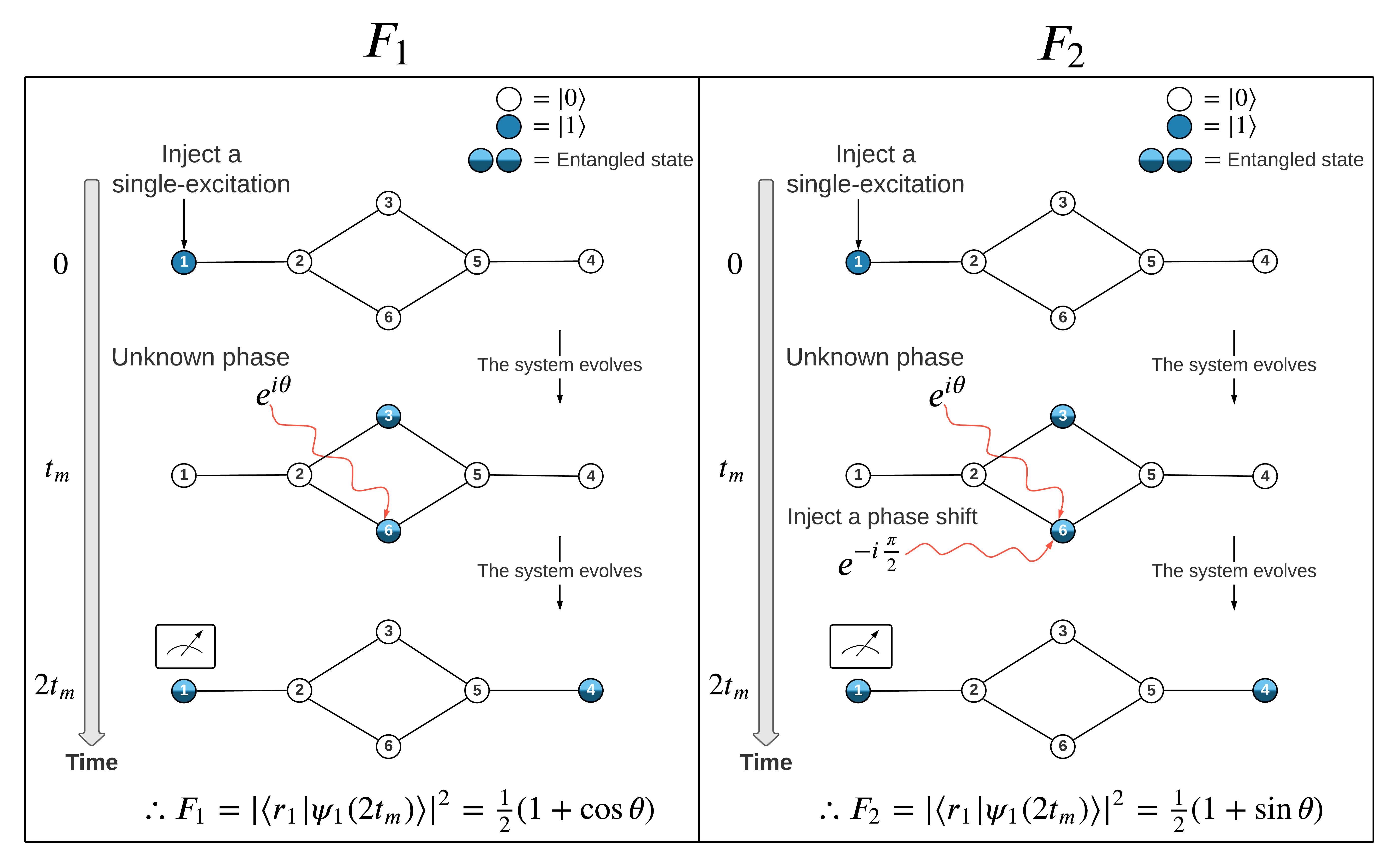}
\caption{Demonstration of two separate experiments used to obtain $F_1$ (left) and $F_2$ (right). In both experiments we choose to measure the fidelity against $\ket{r_1}$.}
\label{fig: F1 and F2 experiments}     
\end{figure*}

In order to obtain a fidelity as a function of $\sin \theta$, an additional sudden phase factor of $\pm i$ should be applied to either site 3 or site 6. For example, a known shift of $-\frac{\pi}{2}$  applied to site 6 gives a total sudden phase factor of ($\exp{i(\theta-\frac{\pi}{2})}$). As a result, the fidelity against $|r_1\rangle$  at $2t_m$ can now be written as 
\begin{eqnarray}
\label{F2 eq} 
F_2 &=&\frac{1}{2}(1+\cos{(\theta-\frac{\pi}{2})})
\\
\label{F22 eq}      
&=&\frac{1}{2}(1+\sin{\theta}). 
\end{eqnarray} 
We label the fidelity index here by 2 in order to distinguish between two fidelities. $F_1$ is the fidelity against $\ket{r_1}$ at $2t_m$ when the unknown phase is applied at site 6, whereas $F_2$ is the fidelity against $\ket{r_1}$ at $2t_m$ when the additional shift of $-\frac{\pi}{2}$ is added to the unknown phase, Fig.~\ref{fig: F1 and F2 experiments}.

\subsubsection{Phase Sensor Protocol and Its Robustness}
\begin{figure}[ht]
    \centering
    \includegraphics[width = 0.45\textwidth]{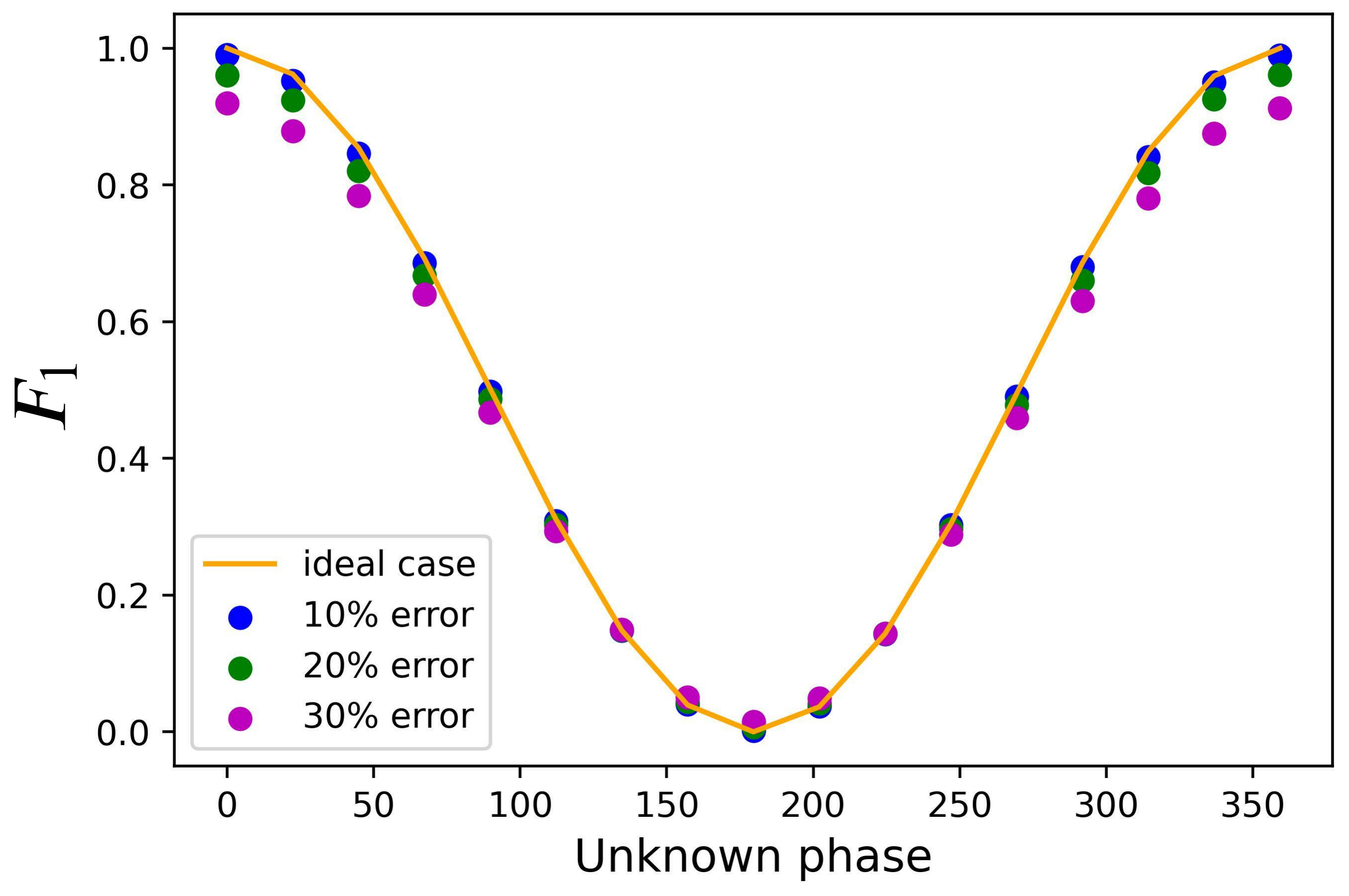} 
    \caption{Fidelity $F_1$ against $\ket{r_1}$ at $2t_M$ in the presence of off-diagonal disorder. The orange line corresponds to the ideal case where there is no error. The dots are the averaged fidelity of 1000 fidelity realisations for each unknown phase with error scaled up to $30\%$. The angles are given in degrees.}
    \label{figgg:1}
\end{figure}

\begin{figure}[ht]
    \centering
    \includegraphics[width = 0.45\textwidth]{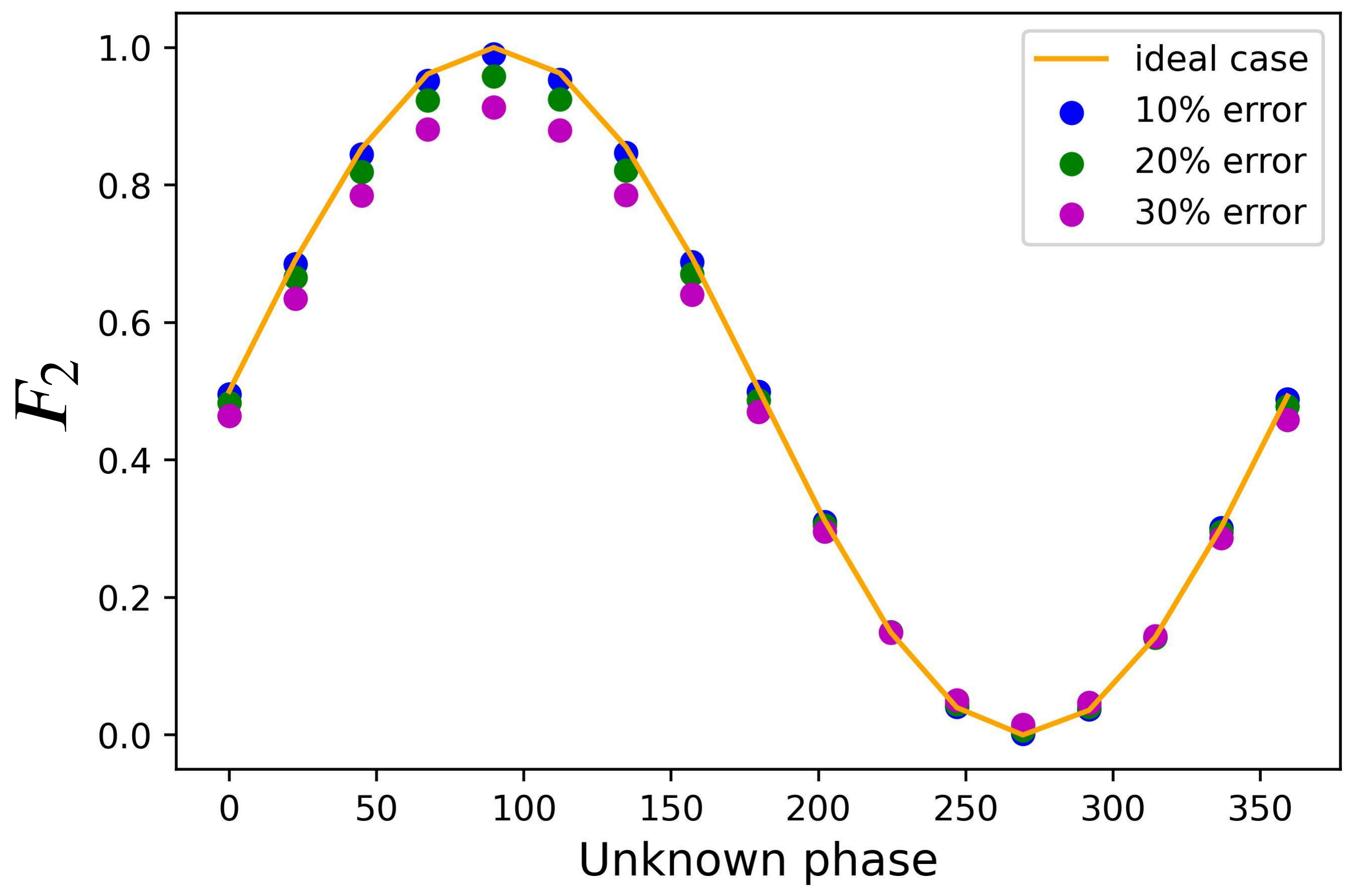} 
    \caption{Fidelity $F_2$ against $\ket{r_1}$ at $2t_M$ in the presence of off-diagonal disorder. The orange line corresponds to the ideal case where there is no error. The dots are the averaged fidelity of 1000 fidelity realisations for each unknown phase with error scaled up to $30\%$. The angles are given in degrees.}
    \label{figgg:2}
\end{figure}

Here, we discuss details of the retrieval protocol and results in the presence of imperfections. Both fidelities $F_1$ and $F_2$ are needed for this protocol, so the injection of the unknown phase must be repeatable.
In practical implementations there will be some disorder, or imperfections, in the system and so we will consider the performance of the phase-sensing protocol in the presence of disorder in the system.  
Results for the fidelities $F_1$ and $F_2$ in the presence of {\it off-diagonal} disorder are presented in Fig.\eqref{figgg:1} and Fig.\eqref{figgg:2}. Results are for a set of unknown phases ranging from 0 to $2\pi$, and are averaged over 1000 realisations.
It is apparent that the fidelity is more sensitive to off-diagonal disorder when it is approaching unity, see Fig.\eqref{figgg:1} and Fig.\eqref{figgg:2}. The {\it diagonal} disorder on the other hand has very weak effect on the fidelity, see appendix \ref{appendix C}.

The phase-sensing protocol given below is designed to account for the more damaging type of disorder, which is the off-diagonal disorder. The distribution of the random number error considered here is the Gaussian distribution. 

It is clear from Fig.\eqref{figgg:1} and Fig.\eqref{figgg:2} that the fidelities approaching unity are reduced in the presence of error, from the actual unit fidelity in the ideal case. As a result, the angles retrieved from these high fidelities will suffer greater error and deviate more from their actual values. To overcome this problem, we propose a flexible protocol based on the behaviour of $F_1$ and $F_2$ for different unknown phases in the presence of error.


Our phase-sensing protocol, that includes error-mitigation, can be understood by examining the behaviour of $F_1$ and $F_2$ in Fig.\eqref{figgg:1} and Fig.\eqref{figgg:2}. This shows that, for a given unknown phase, when $F_1$ is more affected by the error, $F_2$ is less affected, and vice versa. Therefore, our protocol uses both $F_1$ and $F_2$ to obtain two angles $\theta_1$ and $\theta_2$, respectively. Then, we choose the angle with less standard deviation as the best estimate to the unknown angle.

For a given data set \{$F_1$,$F_2$\} associated to an unknown phase, the protocol works as follows: \\
\begin{enumerate}
    \item Use $F_1$ to find $\theta_1$ such that $\theta_1=\cos^{-1}({2F_1-1})$  \\         
    Then, use $F_2$ to determine in which range the theta is: \\
    if $F_2$ $\geq$ 0.5, then $0\leq\theta_1\leq\pi$ \\
    if $F_2 <$ 0.5, then $\pi\leq\theta_1\leq2\pi$ \\ 
    \item Use $F_2$ to find $\theta_2$ such that $\theta_2=\sin^{-1}({2F_2-1})$  \\ 
    Then, use $F_1$ to determine in which range the theta is: \\
    if $F_1$ $\geq$ 0.5, then $-\frac{\pi}{2}\leq\theta_2\leq\frac{\pi}{2}$ \\
    if $F_1 <$ 0.5, then $\frac{\pi}{2}\leq\theta_2\leq\frac{3\pi}{2}$ \\ 
\end{enumerate}
   
Step 2 employs the angle range from $-\frac{\pi}{2}$ to $\frac{3\pi}{2}$, because with this range the use of $F_1$ to determine the range of the $F_2$ angle then uniquely distinguishes between just two continuous regions of $\theta_2$. This is necessary in order for the averaging of angles to work, as we now describe. 

If for an unknown phase we have a set of 1000 random realisations of data \\
$\{\{F_1,F_2\}_1,\{F_1,F_2\}_2,\ldots,\{F_1,F_2\}_{1000}\}$ and we use the above protocol for each set of data, we will obtain $\{ \{\theta_1,\theta_2\}_1, \{\theta_1,\theta_2\}_2, \ldots, \{\theta_1,\theta_2\}_{1000} \}$. We then take the average of all $\theta_1$ as well as the separate average of all $\theta_2$. Note that for some unknown phases the averaged $\overline{\theta_2}$ can be negative because of the negative ranges in step 2 and therefore we need to add $2\pi$ to the negative angles to shift them to be in the range $3\pi/2$ to $2\pi$. Each averaged angle $\overline{\theta_1}$ and $\overline{\theta_2}$ corresponds to the unknown phase angle, but one of these will  deviate more from the actual unknown angle, due to the larger error that affects higher fidelities. Therefore, we use the standard deviations of $\overline{\theta_1}$ and $\overline{\theta_2}$ to determine which approach in the protocol (1 or 2) gives a lower standard deviation and so corresponds to the most accurate value of the unknown angle. 

\begin{figure}[ht!]
    \centering
    \includegraphics[width = 0.45\textwidth]{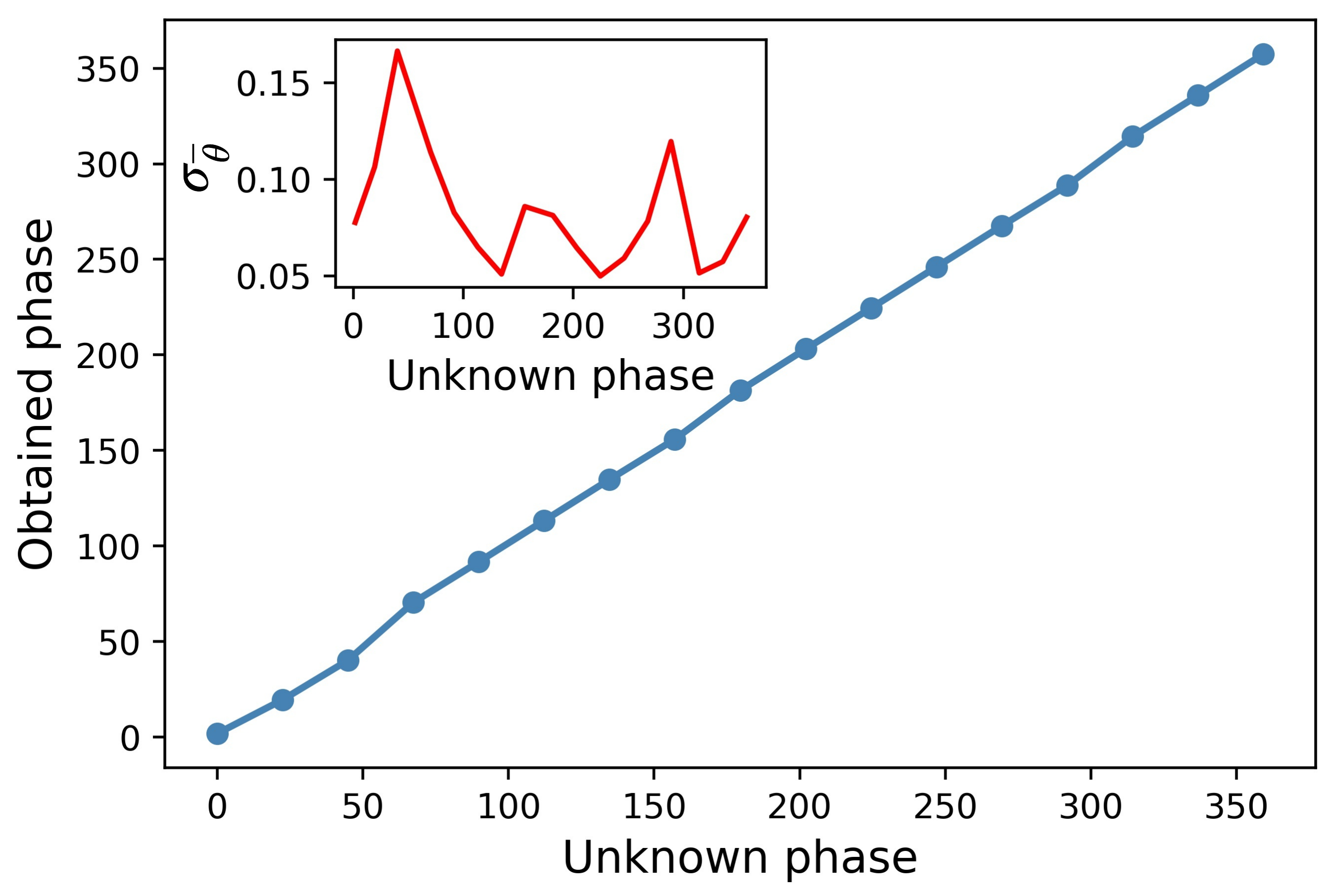} 
    \caption{The obtained angles vs the unknown angles, in degrees. This is in the presence of off-diagonal disorder with error scale $E/J=20\%$ with angles being averaged over 1000 realisations. Inset: the standard deviation, $\sigma_{\overline{\theta}}$, of the mean of the obtained angles. It is clear that the angle around 45\si{\degree} has the highest standard deviation.}
    \label{fig:obtained phases vs unknown phases}
\end{figure} 

The graph generated in Fig.\eqref{fig:obtained phases vs unknown phases} illustrates the performance of our sensing protocol in retrieving unknown phases in the presence of a large off-diagonal disorder with error scale $E/J=20\%$. The inset in the graph shows that the standard deviation of the mean of the obtained phases for an unknown angle of about 45\si{\degree} carries the worst error. With reference to Fig.\eqref{figgg:1} and Fig.\eqref{figgg:2} this is because the first quadrant is where both $F_1$ and $F_2$ lie between 0.5 and 1, and therefore carry non-negligible error. Still even the standard deviation for $\approx45\si{\degree}$ is just a small fraction of one degree, therefore, our SN system can be used as an accurate phase sensor device, with quantified performance, even when the SN device being used contains significant errors.

\section{Conclusion} 
\label{Conclusion}
We have demonstrated the possibility of using ad-hoc unitary transformations to design multifunctional spin-networks starting from uncoupled components of known properties, in our case two short PST chains. The simple spin network system we
obtained can deliver useful tasks for quantum communication,
quantum information, and quantum sensing. It can be used as
a router device for short-distance quantum communication or to
create a maximally entangled state between two chosen network
sites, preparing a resource that could be used in quantum information
processing. Furthermore, we have demonstrated that the
spin network we constructed can be used as a phase sensor device,
that determines with quantified accuracy an unknown phase
applied to a site within the system.

Our modelling investigations of disorder in the system shows that all these applications function with very high fidelity, even in the presence of significant ($\sim$10-20\%) fabrication or slowly varying imperfections. In more detail, modelling investigations of the effect of off-diagonal disorder (the most damaging type of disorder) demonstrate that our SN system retains EOF of $90\%$ or higher for errors up to $E/J=20\%$ and $E/J=10\%$ when the maximally entangled state is collected when it first forms and at $4t_m$, respectively. In the phase sensing protocol, it suffers a maximum error of a fraction of a degree for imperfections up to $E/J=20\%$. High quality devices might be expected to have errors with $E/J<10\%$ or lower, therefore the robustness observed in our SN system suggests this device could support various tasks in the practical application of quantum technology.

In general, off-diagonal disorder is somewhat more damaging
than diagonal disorder, so reduction of off-diagonal disorder
would be the highest experimental priority. However, the system
demonstrates significant robustness against all practical levels of
disorder, as detailed above. In the analysis, we have considered
two types of error distributions, Gaussian and flat distribution.
We conclude that the flat distribution is slightly more damaging
than the Gaussian distribution for very large errors, but else
essentially no dependence on the form of the error distribution
is exhibited.

We have demonstrated that by using ad hoc unitaries to link
building blocks with special properties new behaviors emerge
in the network, which can be exploited for a variety of quantum
information tasks. The application of this method can be
extended well beyond the examples given in this paper. The simplest
case would be to consider longer PST chains, which would
lead, for example, to longer-distance router or entanglement distributors.
Also, here we only considered the single-excitation subspace
of the SN. However, multiple excitation subspaces extend
the range of possible applications of SN systems, one example
being the possibility of engineering unitaries for quantum gates
using the zero-, one-, and two-excitation subspaces. Furthermore,
the method we propose—the engineering of coupled networks
via appropriate unitaries applied to uncoupled spin chains—is
also applicable to higher-excitation subspaces. We plan to explore
this in future work.


\section{Acknowledgment} 
AHA acknowledges support from the WW Smith Fund at the University of York.

\appendix
\section{Eigenvalues and Eigenvectors of the Hamiltonian}
\label{appendix A}
The eigenvalues and eigenvectors of the Hamiltonian $\mathcal{H}$ are given in table \ref{eeH}. They are expressed in the site basis as labelled in Fig.\eqref{fig:6SN}.
\begin{table}[H]
    \centering
    \rowcolors{2}{gray!10}{gray!40}
    \begin{tabular}{cc}
    Eigenvalues & Eigenvectors \\
    \hline
    $\lambda_1=-\sqrt{2}J$ & $\ket{\phi_1}=\frac{1}{2\sqrt{2}}\begin{pmatrix}  
    \sqrt{2} \\
    -2 \\
    1 \\
    0 \\
    0 \\ 
    1 \\
    \end{pmatrix}$ \\
    $\lambda_2=-\sqrt{2}J$ & $\ket{\phi_2}=\frac{1}{2\sqrt{2}}\begin{pmatrix}  
    0 \\
    0 \\
    -1 \\
    -\sqrt{2} \\
    2 \\
    1 \\
    \end{pmatrix}$ \\
    $\lambda_3=\sqrt{2}J$ & $\ket{\phi_3}=\frac{1}{2\sqrt{2}}\begin{pmatrix}  
    \sqrt{2} \\
    2 \\
    1 \\
    0 \\
    0 \\
    1 \\
    \end{pmatrix}$ \\
    $\lambda_4=\sqrt{2}J$ & $\ket{\phi_4}=\frac{1}{2\sqrt{2}}\begin{pmatrix}  
    0 \\
    0 \\
    1 \\
    \sqrt{2} \\
    2 \\
    -1 \\
    \end{pmatrix}$ \\
    $\lambda_5=0$ & $\ket{\phi_5}=\frac{1}{2}\begin{pmatrix}  
    -\sqrt{2} \\
    0 \\
    1 \\
    0 \\
    0 \\
    1 \\
    \end{pmatrix}$ \\
    $\lambda_6=0$ & $\ket{\phi_6}=\frac{1}{2}\begin{pmatrix}  
    0 \\
    0 \\
    -1 \\
    \sqrt{2} \\
    0 \\
    1 \\
    \end{pmatrix}$ \\
    \end{tabular}
    \caption{Eigenvalues (left) and eigenvectors (right) of the Hamiltonian $\mathcal{H}$}
    \label{eeH}
\end{table}

\section{Analytical Calculation of Relevant System Evolution and Related Fidelities}
\label{appendix B}
The SN is first prepared so that all sites have spin down $\ket{00\dots}$. Since the Hamiltonian of our system preserves the number of spin up (down), the system will not evolve unless an excitation is injected to the system. Therefore, when a single-excitation is injected at site 1 at $t=0$
\begin{equation}
\ket{\psi_1(0)}=\ket{r_1}
\end{equation}
the system will start evolving within the single-excitation (single spin-up) subspace. The state of the system at later time can be found by decomposing the initial state $\ket{\psi_1(0)}$ into the eigenvectors of the Hamiltonian (table \ref{eeH}) using the following decomposition equation
\begin{equation}
\label{equation17}  	
\ket{\psi_1(t)} = \sum_{j=1}^{N} \braket{\phi_j\vert\psi_1(0)}e^{-i\lambda_jt} \ket{\phi_j},   
\end{equation}
where $\ket{\phi_j}$ are the eigenvectors, and $\lambda_j$ are the eigenvalues.
Therefore, the state of the system at $t_m$ will be 
\begin{equation} 
\label{appendix B. Eq3&6}      
\ket{\psi_1(t_m)}= -\frac{1}{\sqrt{2}}(\ket{r_3}+\ket{r_6}).
\end{equation}

If now an unknown phase $e^{i\theta}$ is instantaneously applied at site 6, the state of the system becomes
\begin{equation} 
\label{appendix B. Eq3& pf_6}      
\ket{\psi_1(t_m)}_{\theta} = -\frac{1}{\sqrt{2}}(\ket{r_3} + e^{i\theta}\ket{r_6}).
\end{equation}
By decomposing the state Eq.~(\ref{appendix B. Eq3& pf_6}) into the eigenvectors of the Hamiltonian, using Eq.~(\ref{equation17}), and then evolving the system for an additional time $t=t_m$, the state of the system at $2t_m$ can be found to be 
\begin{equation}   
\label{appendix B equation24}   
\ket{\psi_1(2t_m)} = \frac{1+e^{i\theta}}{2}\ket{r_1} +    \frac{1-e^{i\theta}}{2}\ket{r_4}. 
\end{equation}     
Therefore, Eq.~(\ref{appendix B equation24}) determines the occupation of site 1 and site 4 depending on the unknown phase $\theta$. 

The fidelity against $\ket{r_1}$  at $2t_m$ is then 
\begin{equation}
\label{F1 eq} 
\begin{split}
|\braket{r_1|\psi_1(2t_m)}|^2 & = |\frac{1+e^{i\theta}}{2}\braket{r_1|r_1}+\frac{1-e^{i\theta}}{2}\braket{r_1|r_4}|^2\\
& = |\frac{1+e^{i\theta}}{2}|^2 = |e^{i\frac{\theta}{2}}(\frac{e^{-i\frac{\theta}{2}}+e^{i\frac{\theta}{2}}}{2})|^2 \\
& = \cos^2\frac{\theta}{2} = \frac{1+\cos(\theta)}{2} \\   
& =\frac{1}{2}(1+\cos{\theta})=F_1.
\end{split}        
\end{equation}
We now have the fidelity against $\ket{r_1}$ at $2t_m$ as a function of $\theta$ which can be used in our sensing protocol by measuring the fidelity against $\ket{r_1}$. However, if we choose to measure the fidelity against $\ket{r_4}$ instead, then the sensing protocol will still work with a slight modification as the fidelity against $\ket{r_4}$ differs from the fidelity against $\ket{r_1}$, and can be calculated in a similar way. \\

\section{Robustness of the Phase-Sensing Protocol in the Presence of Diagonal Disorder}
\label{appendix C}
\begin{figure}[ht!]
    \centering
    \includegraphics[width = 0.45\textwidth]{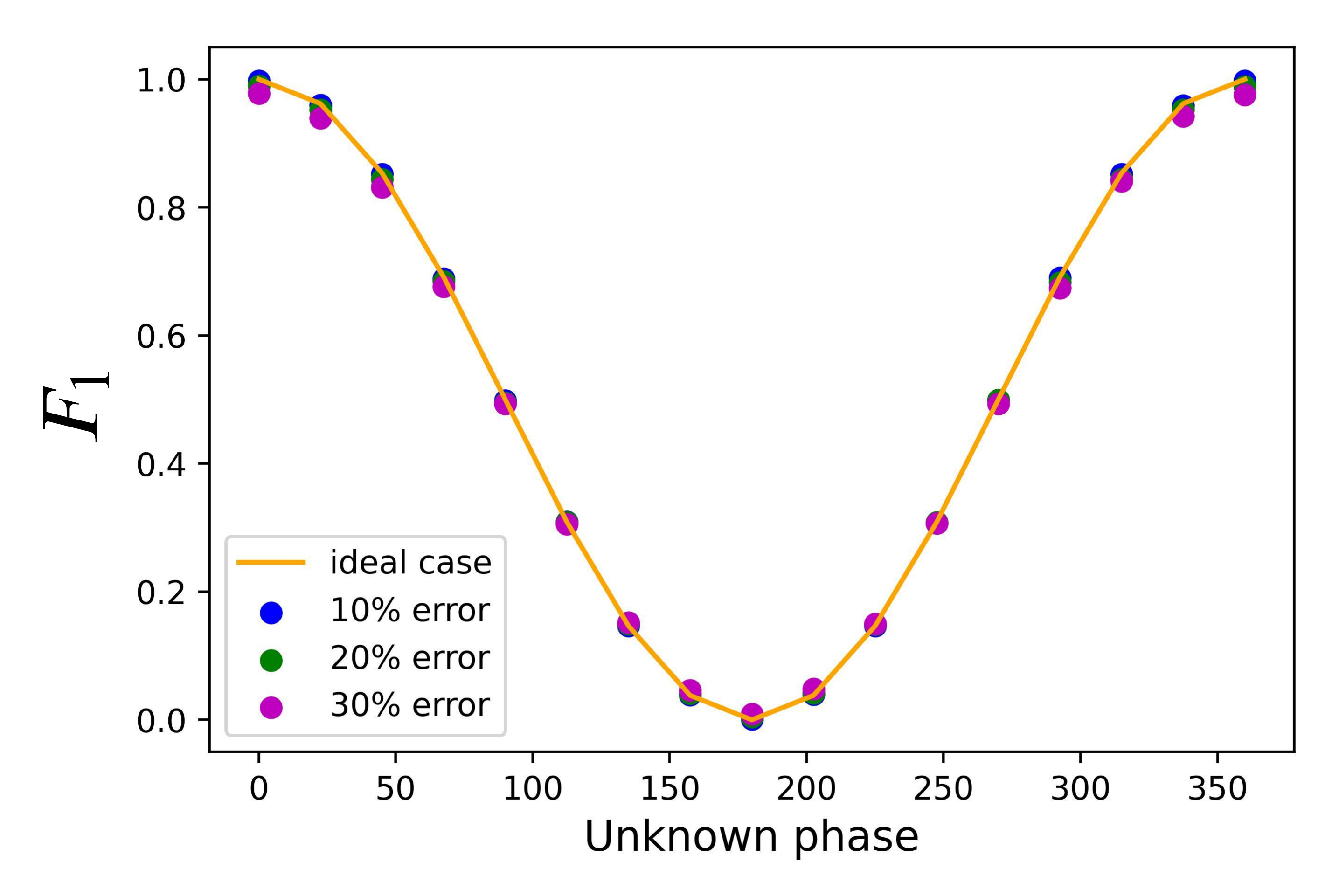} 
    \caption{Fidelity $F_1$ against $\ket{r_1}$ at $2t_M$ in the presence of diagonal disorder. The orange line corresponds to the ideal case where there is no error. The dots are the averaged fidelity of 1000 fidelity realisations for each unknown phase with error scaled up to $30\%$. The angles are given in degrees.}
    \label{fig1:appendix C}
\end{figure}

\begin{figure}[ht!] 
    \centering
    \includegraphics[width = 0.45\textwidth]{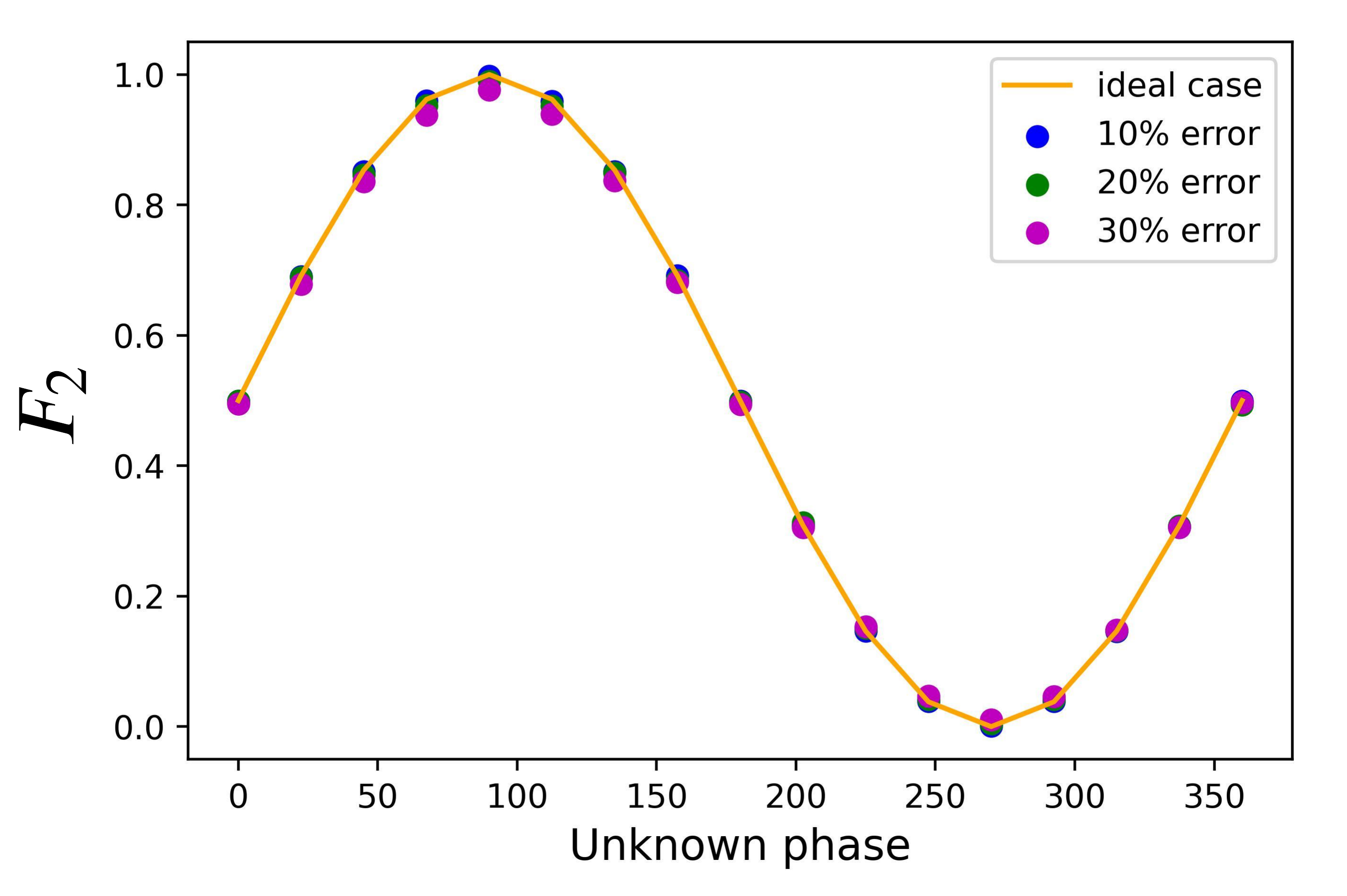} 
    \caption{Fidelity $F_2$ against $\ket{r_1}$ at $2t_M$ in the presence of diagonal disorder. The orange line corresponds to the ideal case where there is no error. The dots are the averaged fidelity of 1000 fidelity realisations for each unknown phase with error scaled up to $30\%$. The angles are given in degrees.}
    \label{fig2:appendix C}
\end{figure}

The fidelities $F_1$ and $F_2$ are very robust against diagonal disorder, see Fig.\ref{fig1:appendix C} and Fig.\ref{fig2:appendix C}. . 

\clearpage

\bibliographystyle{unsrt}   
\bibliography{reff} 

\end{document}